\documentclass[preprint,12pt]{elsarticle}

\usepackage{amssymb}
\usepackage{amsfonts}
\usepackage{epstopdf}
\usepackage{graphicx}
\usepackage{stmaryrd}
\usepackage{amssymb,amsmath}
\usepackage{multirow,url}
\usepackage{color}
\usepackage{hyperref}
\usepackage{enumitem}

\ifodd 0
\newcommand{\com}[1]{\textbf{\color{red} (COMMENT: #1)}} %comment of the text
\newcommand{\comg}[1]{\textbf{\color{green} (COMMENT: #1)}}
\newcommand{\response}[1]{\textbf{\color{magenta} (RESPONSE: #1)}} %response to comment
\else

\newcommand{\com}[1]{}
\newcommand{\comg}[1]{}
\newcommand{\response}[1]{}
\fi
\ifodd 0
\newcommand{\referred}[1]{\textcolor{red}{RefPaper: #1}} %comment of the text
\else
\newcommand{\referred}[1]{}
\fi
\journal{Physical Communication}

\begin{document}

\begin{frontmatter}


% \author{Name\corref{cor1}\fnref{label2}}
% \address{Address\fnref{label3}}

\title{Implementation of Physical-layer Network Coding}

 \author[label1]{Lu Lu} \ead{ll007@ie.cuhk.edu.hk}
 \author[label1]{Taotao Wang} \ead{postman511@gmail.com}
 \author[label1]{Soung Chang Liew} \ead{soung@ie.cuhk.edu.hk}
 \author[label1,label2]{Shengli Zhang} \ead{zsl@szu.edu.cn}
 \address[label1]{Dept. of Information Engineering, The Chinese University of Hong Kong, Hong Kong}
 \address[label2]{Dept. of Communication Engineering, Shenzhen University, China}

%\author{}
%
%\address{}

\begin{abstract}
This paper presents the first implementation of a two-way relay network based on the principle of physical-layer network coding. To date, only a simplified version of physical-layer network coding (PNC) method, called analog network coding (ANC), has been successfully implemented. The advantage of ANC is that it is simple to implement; the disadvantage, on the other hand, is that the relay amplifies the noise along with the signal before forwarding the signal. PNC systems in which the relay performs XOR or other denoising PNC mappings of the received signal have the potential for significantly better performance. However, the implementation of such PNC systems poses many challenges. For example, the relay must be able to deal with symbol and carrier-phase asynchronies of the simultaneous signals received from the two end nodes, and the relay must perform channel estimation before detecting the signals. We investigate a PNC implementation in the frequency domain, referred to as FPNC, to tackle these challenges. FPNC is based on OFDM. In FPNC, XOR mapping is performed on the OFDM samples in each subcarrier rather than on the samples in the time domain. We implement FPNC on the universal soft radio peripheral (USRP) platform. Our implementation requires only moderate modifications of the packet preamble design of 802.11a/g OFDM PHY. With the help of the cyclic prefix (CP) in OFDM, symbol asynchrony and the multi-path fading effects can be dealt with in a similar fashion. Our experimental results show that symbol-synchronous and symbol-asynchronous FPNC have essentially the same BER performance, for both channel-coded and unchannel-coded FPNC.
\end{abstract}


\end{frontmatter}

% \linenumbers

\section{Introduction}
In this paper, we present the first implementation of physical-layer network coding (PNC) on the software radio platform. We believe this prototyping effort moves the concept of PNC a step toward reality. Our implementation work also exposes and raises some interesting issues for further research.
%\com{Will re-examine this last sentence later. Other than CFO, are there other interesting issues exposed/raised by us? How about MAC protocol design? Implementation of time-domain PNC? Other desnoising PNC?}

\begin{figure}[h]
\centering
\includegraphics[width=0.7\textwidth]{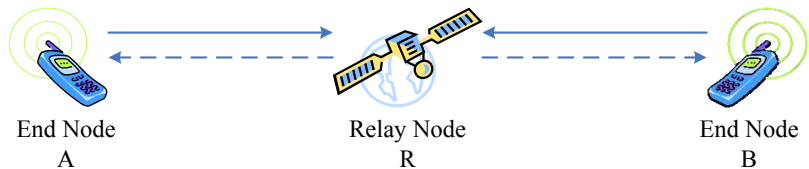}
\caption{System model for physical-layer network coding.} \label{fig:SystemModel}
\end{figure}

PNC, first proposed in \referred{PNC06}\cite{PNC06}, is a subfield of network coding \referred{AhlswIT00}\cite{AhlswIT00} that is attracting much attention recently. The simplest system in which PNC can be applied is the two-way relay channel (TWRC), in which two end nodes $A$ and $B$ exchange information with the help of a relay node $R$ in the middle, as illustrated in Fig. \ref{fig:SystemModel}. Compared with the conventional relay system, PNC could double the throughput of TWRC by reducing the needed time slots for the exchange of one packet from four to two \referred{PNC06}\cite{PNC06}. In PNC, in the first time slot, end nodes $A$ and $B$ send signals simultaneously to relay $R$; in the second phase, relay $R$ processes the superimposed signals and maps them to a network-coded packet for broadcast back to the end nodes. From the network-coded packet, each end node then makes use of its self information to extract the packet from the other end node \referred{PNC06,ShengliGlobecom08, PNCSurvey}\cite{PNC06,ShengliGlobecom08,PNCSurvey}.

Prior to this paper, only a simplified version of PNC, called analog network coding (ANC) \referred{KattiANC07}\cite{KattiANC07}, has been successfully implemented. The advantage of ANC is that it is simple to implement; the disadvantage, on the other hand, is that the relay amplifies the noise along with the signal before forwarding the signal, causing error propagation.

To our best knowledge, the implementation of the original PNC based on XOR mapping as in \referred{PNC06}\cite{PNC06} has not been demonstrated, even though it could have significantly better performance. A reason is that the implementation of XOR PNC poses a number of challenges. For example, the relay must be able to deal with symbol and carrier-phase asynchronies of the simultaneous signals received from the two end nodes, and the relay must perform channel estimation before detecting the signals.

This paper presents a PNC implementation in the frequency domain, referred to as FPNC, to tackle these challenges. In particular, FPNC is based on OFDM, and XOR mapping is performed on OFDM samples in each subcarrier rather than the samples in the time domain. We implement FPNC on the universal soft radio peripheral (USRP) platform. Our implementation requires only moderate modifications of the packet preamble design of 802.11a/g OFDM PHY. With the help the cyclic prefix (CP) in OFDM, symbol asynchrony and the multi-path fading effects can be dealt with in a similar fashion. Our experimental results show that symbol-synchronous and symbol-asynchronous FPNC have nearly the same BER performance, for both channel-coded and unchannel-coded FPNC.

\setcounter{secnumdepth}{-1}% cancel the chapter number for the Introduction
\subsection{\textbf{Challenges}}
In the following, we briefly overview the challenges of PNC, and the implementation approaches taken by us to tackle them:

\subsubsection{\texttt{Asynchrony}}
There are two possible implementations for PNC: synchronous PNC and asynchronous PNC. In synchronous PNC, end nodes $A$ and $B$ have the uplink channel state information (CSI). They perform precoding and synchronize their transmissions so that their signals arrive at relay R with symbol and carrier-phase alignments. For high-speed transmission, such tight synchronization is challenging; in addition, timely collection of CSI is difficult in fast fading scenarios.

Asynchronous PNC is less demanding. It does not require the two end nodes to tightly synchronize and precode their transmissions. In particular, knowledge of the uplink CSI is not needed at the two end nodes. The simplicity at the end nodes comes with a cost. Without precoding and synchronization of the two end nodes, their signals may arrive at the relay with symbol and carrier-phase misalignments. A key issue in asynchronous PNC is how to deal with such signal asynchrony at the relay \referred{SyncPNC06, UPNCICC11}\cite{SyncPNC06, UPNCICC11}.

\begin{figure}[h]
\centering
\includegraphics[width=1\textwidth]{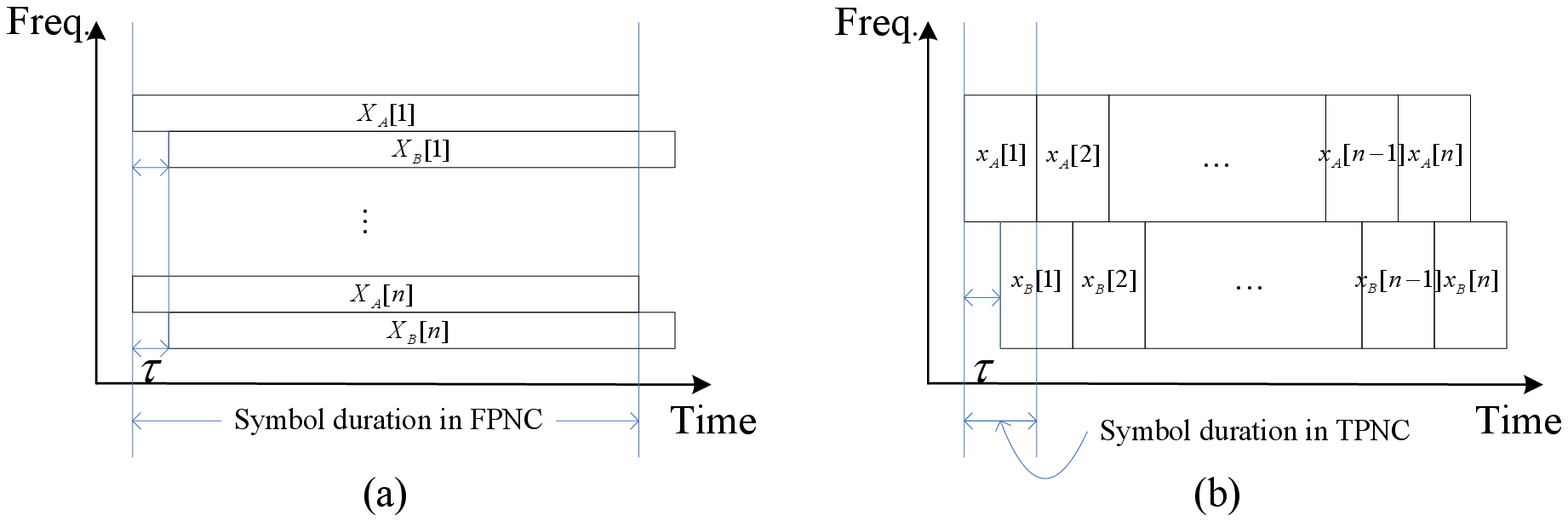}
\caption{PNC with time asynchrony: (a) frequency-domain physical-layer network coding (FPNC); (b) time-domain physical-layer network coding (TPNC).}\label{fig:FreqvsTimePNC}
\end{figure}

This paper focuses on the implementation of asynchronous PNC. To deal with asynchrony, our FPNC implementation makes use of OFDM to lengthen the symbol duration within each subcarrier. Then, independent XOR PNC mapping is performed within each subcarrier. OFDM splits a high-rate data stream into a number of lower-rate streams over a number of subcarriers. Thanks to the larger symbol duration within each subcarrier, the relative amount of dispersion caused by the multipath delay spread is decreased. The OFDM symbols of the two end nodes become more aligned with respect to the total symbol duration, as illustrated in Fig. \ref{fig:FreqvsTimePNC}. In particular, if the relative symbol delay is within the length of the CP, the time-domain misaligned samples will become \emph{aligned} in the frequency domain after DFT is applied. This property will be further elaborated later in Section 2.

\subsubsection{\texttt{Channel Estimation}}
For good performance of asynchronous PNC, the relay must have the knowledge of the uplink CSI. This has been the assumption in many prior works on PNC (e.g., \referred{PNC06, ZorziSPAWC09}\cite{PNC06, ZorziSPAWC09}).  This means that in implementation, the relay will need to estimate the channel gains. Most channel estimation techniques for the OFDM system assume point-to-point communication in which only one channel needs to be estimated. In PNC, the relay needs to estimate two channels based on simultaneous reception of signals (and preambles) from the two end nodes. This poses the following two problems in PNC that do not exist in point-to-point communication:

\begin{itemize}
\item Channel estimation in a point-to-point OFDM system (e.g., 802.11 \referred{dot11std07}\cite{dot11std07}) is generally facilitated by training symbols and pilots in the transmitted signal. If used unaltered in the PNC system, the training symbols and pilots from the two end nodes may overlap at the relay, complicating the task of channel estimation. In our implementation, we solve this problem by assigning orthogonal training symbols and pilots to the end nodes. The details will be given in Section 4.

\item It is well known that carrier frequency offset (CFO) between the transmitter and the receiver can cause inter-subcarrier interference (ICI) if left uncorrected. In a point-to-point system, CFO can be estimated and compensated for. In PNC, we have two CFOs at the relay, one with respect to each end node. Even if the two CFOs can be estimated perfectly, their effects cannot be both compensated for totally; the total elimination of the ICI of one end node will inevitably lead to a larger ICI for the other end node. To strike a balance, our solution is to compensate for the mean of the two CFOs (i.e., compensate for ${\rm{(CFO}}_A  - {\rm{CFO}}_B )/2$). The details will be elaborated in Section 4.

\end{itemize}

\subsubsection{\texttt{Joint Channel Decoding and Network Coding}}
For reliable communication in a practical PNC system, channel coding needs to be incorporated. This paper considers link-by-link channel-coded PNC, in which the relay maps the overlapped channel-coded symbols of the two end nodes \referred{PNCSurvey,ShengliJSAC09}\cite{PNCSurvey,ShengliJSAC09} to the XOR of the source symbols \footnote{This process is called Channel-decoding-Network-Coding (CNC) in \referred{ShengliJSAC09}\cite{ShengliJSAC09} because it does two things: channel decoding and network coding. Unlike the traditional multiuser detection (MUD) in which the goal is to recover the individual source information from the two end nodes, CNC aims to recover the XOR of the source information during the channel decoding process. CNC is a component in link-by-link channel-coded PNC critical for its performance \referred{PNCSurvey,ShengliJSAC09}\cite{PNCSurvey,ShengliJSAC09}.}; after that, the relay channel-encodes the XOR source symbols to channel-coded symbols for forwarding to the end nodes. Such a link-by-link channel-coded PNC system has better performance than an end-to-end channel-coded PNC system \referred{PNCSurvey,ShengliJSAC09}\cite{PNCSurvey,ShengliJSAC09}.

In our FPNC design, we adopt the convolution code as defined in the 802.11 a/g standard. The relay first maps the overlapped channel-coded symbols to their XOR on a symbol-by-symbol basis. After that it cleans up the noise by (i) channel-decoding the XOR channel-coded symbols to the XOR source symbols, and then (ii) re-channel-coding the XOR source symbols to the XOR channel-coded symbols for forwarding to the two end nodes.

\vspace{0.15in}
The remainder of this paper is organized as follows: Section 2 details the delay asynchrony model of this paper. Section 3 presents the FPNC frame format design. Section 4 addresses the key implementation challenges. Experimental results are given in Section 5. Finally, Section 6 concludes this paper.

%\begin{figure}[h]
%\centering
%\includegraphics[width=1\textwidth]{test.eps}
%\caption{PNC with time asynchrony: (a) frequency-domain physical-layer network coding (FPNC); (b) time-domain physical-layer network coding (TPNC).}\label{fig:FTPNC}
%\end{figure}
\setcounter{secnumdepth}{3} %set the number of chapter to be 3
\section{Effect of Delay Asynchrony in Frequency Domain}
In asynchronous PNC, symbols of the two end nodes may arrive at the relay misaligned. We mentioned in the introduction that if the relative symbol delay is within the length of the CP in FPNC, then the time-domain misaligned samples will become \emph{aligned} in the frequency domain after DFT is applied. This section is devoted to the mathematical derivation of this result. Here, we will derive a more general result that takes into account multi-path channels as well.

\subsection{Effective Discrete-time Channel Gains}
We consider the following multi-path channel model. Suppose that there are $M_A$ paths from node $A$ to relay $R$ with delays $\tau _A^0  < \tau _A^1  <  \cdots  < \tau _A^{M_A -1 }$ and corresponding channel gains $\alpha_A^0 , \alpha_A^1 , \ldots , \alpha_A^{M_A -1 }$. The channel impulse response of $A$ is $g_A (t) = \sum\limits_{i = 0}^{M_A -1} {\alpha _A^i \delta (t - \tau _A^i )}$. Similarly, there are $M_B$ paths from node $B$ to relay $R$ with delays $\tau _B^0 < \tau _B^1  < \cdots < \tau _B^{M_B - 1}$ and channel gains $\alpha _B^0, \alpha _B^1 , \ldots, \alpha _B^{M_B - 1}$, with  channel impulse response $g_B (t) = \sum\limits_{i = 0}^{M_B -1} {\alpha _B^i \delta (t - \tau _B^i )}$. Without loss of generality, we assume that frame $A$ arrives earlier than frame $B$: specifically, $\tau _A^0  \le \tau _B^0$. \emph{Note that our model allows for the case where nodes $A$ and $B$ do not exactly transmit at the same time.} If one node transmits slightly later than the other, we could simply add the lag time to all the path delays of that node. We assume that the net effect is such that the signal of $A$ arrives earlier than the signal of $B$, whether this is due to earlier transmission or shorter path delay of $A$ .

\begin{figure}[h]
\centering
\includegraphics[width=1\textwidth]{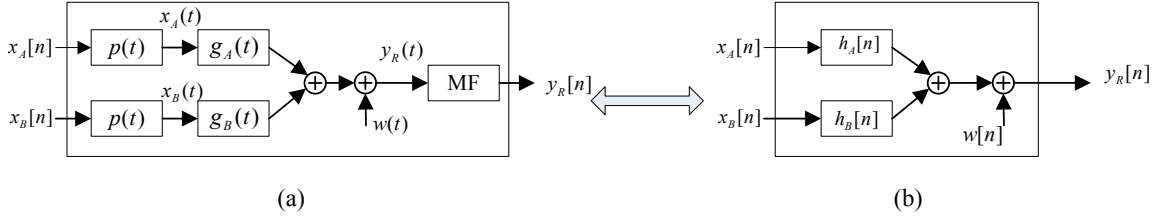}
\caption{(a) Continuous-time channel model for PNC, in which $x_A[n]$ and $x_B[n]$ are the time domain source samples; $y[n]$ is the time domain received samples; $g_A(t)$ and $g_B(t)$ are the wireless multipath channel gains; $p(t)$ is the pulse shaping function; $w(t)$ is the receiver noise; and MF is the matched filter and sampler at the relay node. (b) Equivalent discrete-time channel model for PNC, in which $h_A[n]$ and $h_B[n]$ denote the equivalent discrete time channel impose response (i.e., effective discrete time channel gains), and $w[n]$ is the equivalent discrete-time noise term.}
\label{fig:DiscTimeChannel}
\end{figure}

We first derive the effective discrete-time channel gains for the uplink in FPNC. As shown in Fig. \ref{fig:DiscTimeChannel}, the discrete-time channel gains capture not just the continuous-time channel gains, but also the operations performed by pulse shaping and matched-filtering-and-sampling. Let us assume that the pulse shaping function $p(t)$ is of finite length: specifically, we assume $p(t) = 0$ for $t \le 0$ and $t \ge T_P $. The continuous-time baseband signal fed into the continuous-time channel is $x_A (t) = \sum\limits_{n =  - \infty }^\infty  {x_A [n]p(t - nT)} $. The time domain received signal if only node $A$ transmits is
\begin{align}
y_A (t) = x_A (t)*g_A (t) + w(t) = \sum\limits_{n =  - \infty }^\infty  {\sum\limits_{i = 0}^{M_A - 1} {\alpha _A^i x_A [n]p(t - \tau _A^i  - nT)} }  + w(t),
\label{equ:Oneuser}
\end{align}
where $w(t)$ is the noise, assumed to be AWGN. Matched-Filtering (MF) and sampling are then performed on (\ref{equ:Oneuser}), by sampling at the first multipath channel tap of the uplink channel between node $A$ and relay $R$, to get the received samples
%\begin{align}
%\begin{array}{l}
% y_A [m] = \int\limits_{ - \infty }^\infty  {y(t)p(t - \tau _A^0  + T_P  - mT)} dt \\
%  = \sum\limits_{n =  - \infty }^\infty  {x_A [n]\left\{ {\int\limits_{ - \infty }^\infty  {\sum\limits_{i = 0}^{M_A - 1} {\alpha _A^i p(t - \tau _A^i  - nT)} p(t - \tau _A^0  + T_P  - mT)} dt} \right\}}  + w[m] \\
%  = \sum\limits_{n =  - \infty }^\infty  {x_A [n]h_A [m - n]}  + w[m],
% \end{array}
%\label{equ:Oneuser1}
%\end{align}
\begin{align}
 y_A [m] &= \int\limits_{ - \infty }^\infty  {y(t)p(t - \tau _A^0  + T_P  - mT)} dt \nonumber\\
  &= \sum\limits_{n =  - \infty }^\infty  {x_A [n]\left\{ {\int\limits_{ - \infty }^\infty  {\sum\limits_{i = 0}^{M_A - 1} {\alpha _A^i p(t - \tau _A^i  - nT)} p(t - \tau _A^0  + T_P  - mT)} dt} \right\}}  + w[m] \nonumber\\
  &= \sum\limits_{n =  - \infty }^\infty  {x_A [n]h_A [m - n]}  + w[m],
\label{equ:Oneuser1}
\end{align}
where $w[m] = \int\limits_{ - \infty }^\infty  {w(t)p(t - \tau _A^0  + T_P  - mT)} dt$. We see that the effective discrete-time channel of $A$ is such that $h_A [m - n] = \int\limits_{ - \infty }^\infty  {\sum\limits_{i = 0}^{M_A - 1} {\alpha _A^i p(t - \tau _A^i  - nT)} p(t - \tau _A^1  + T_P  - mT)} dt$.

Note that $p(t - \tau _A^i  - nT)p(t - \tau _A^0  + T_P  - mT) = 0$ if $\left| {\tau _A^i  - \tau _A^0  + T_P  - (m - n)T} \right| \ge T_P $. In other words, $h_A [m - n] = 0$ for $(m - n)T \ge \tau _A^{M_A - 1}  - \tau _A^0  + 2T_P $ and $(m - n)T \le 0$. Define $D_A  = \left\lceil {\left( {\tau _A^{M_A - 1}  - \tau _A^0  + 2T_P } \right)/T} \right\rceil $.

Let us now consider what if both end nodes transmit. The received signal at the relay node is
\begin{align}
y_R (t) = x_A (t)*g_A (t) + x_B (t)*g_B (t) + w(t).
\label{equ:System}
\end{align}
Sticking to the above MF that is defined with respect to first path delay of $A$, we have
\begin{align}
y_R [n] = x_A [n]*h_A [n] + x_B [n]*h_B [n] + w[n],
\label{equ:System1}
\end{align}
where $h_A [n] = 0$ for $n < 0$ and $n \ge D_A  \buildrel \Delta \over = \left\lceil {\left( {\tau _A^{M_A - 1}  - \tau _A^0  + 2T_P } \right)/T} \right\rceil$, and $h_B [n] = 0$ for $n < \left\lceil {\left( {\tau _B^{0}  - \tau _A^0  + 2T_P } \right)/T} \right\rceil$ and $n \ge D_B  \buildrel \Delta \over = \left\lceil {\left( {\tau _B^{M_B - 1}  - \tau _A^0  + 2T_P } \right)/T} \right\rceil$.

\com{  probably for $n <= $ some value larger than 0 ¡­ probably $\frac{{\left\lceil {\tau _B^1  - \tau _A^1  + 2T_P } \right\rceil }}{T}$ or the floor of something like that. Need to go back to the inequality above¡­. }

%$D_A  = \left\lceil {\left( {\tau _A^{M_A - 1 }  - \tau _A^0  + 2T_P } \right)/T} \right\rceil $.

\begin{figure}[h]
\centering
\includegraphics[width=0.6\textwidth]{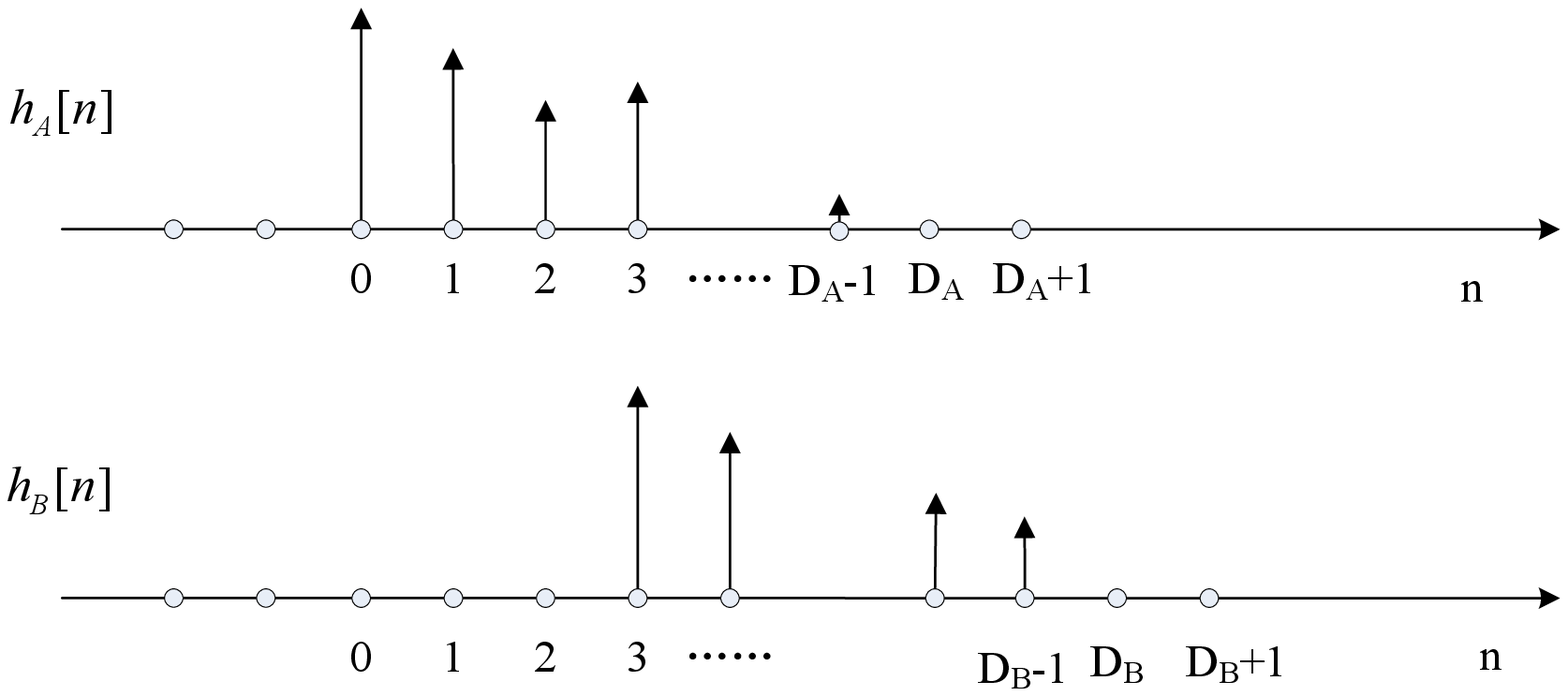}
\caption{Example of delay spread in FPNC.}\label{fig:DelaySpread}
\end{figure}

\subsection{Delay-Spread-Within-CP Requirement}
\com{Revise the whole section by adding CP}
The delay spread of node $A$ is $D_A$, and the delay spread of node $B$, with respect to time $n=0$, is $D_B$. We define the delay spread of the PNC system as (i.e., it combines the delay spreads of $A$ and $B$ into a potentially larger delay spread, as illustrated in Fig. DelaySpread)
%\begin{align}
%\rm{delay}\ \rm{spread} = \max \left[ D_A, D_B  \right].
%\label{equ:DelaySpread}
%\end{align}
\begin{align}
\text{delay spread} = \max \left[ D_A, D_B  \right].
\label{equ:DelaySpread}
\end{align}
The above derivation is general and do not have any requirement on the modulation. This subsection, we will present the OFDM modulated PNC system. In particular, we will present the ``Delay-Spread-Within-CP Requirement'' for FPNC. That is, we combine the Cyclic Prefix (CP) and Discrete Fourier Transform (DFT) to show that the time-domain symbol asynchrony of FPNC disappears in the frequency domain, when the uplink frames satisfy the Delay-Spread-Within-CP Requirement.

First, let $H[k]$ be the $N$-point DFT of $h[n]$ given by \referred{OppenheimDTSPBook}\cite{OppenheimDTSPBook} (Note that, in the following derivation, we assume the subcarrier indices start from $0$, i.e, $k=0,\cdots,N-1$)
\begin{align}
H[k] = DFT\{ h[n]\}  = \sum\limits_{n = 0}^{N-1} {h[n]e^{ - j\frac{{2\pi nk}}{N}} },\ \ \ \ \ \  0 \le k \le N-1.
\label{equ:DFT}
\end{align}
The $N$-point circular convolution of $x_n$ and $h_n$ is written as
\begin{align}
y[n] = x[n] \otimes_N h[n] = \sum\limits_{k = 0}^{N - 1} {h[k ]x[n - k]_N },
\label{equ:CircularConv}
\end{align}
where $[n - k]_N$ denotes $[n - k]$ modulo $N$. In other words, $x[n - k]_N $ is a periodic version of $x[n - k]$ with period $N$. From the definition of DFT, circular convolution in time leads to multiplication in the frequency \referred{OppenheimDTSPBook}\cite{OppenheimDTSPBook}:
\begin{align}
DFT\{ x[n] \otimes_N h[n]\} = X[k]H[k], \ \ \ \ \ \ 0 \le k \le N-1.
\label{equ:CircularConv1}
\end{align}
The channel output, as in (\ref{equ:System1}), however, is not a circular convolution but a linear convolution. \emph{The linear convolution between the channel input and impulse response can be turned into a circular convolution by adding a special prefix to the input called a cyclic prefix (CP)} \referred{GoldsmithWCBook}\cite{GoldsmithWCBook}.

For FPNC, let $H_A[k]$ and $H_B[k]$ denote the frequency responses of the discrete-time channels, and let $C$ denote the length of the CP. One OFDM symbol duration is then $N+C$. The CP for $x_A[n]$ is defined as ${x_A[N-C], \ldots, x_A[N-1]}$: it consists of the last $C$ values of the $x_A[n]$ sequence. For each input sequence of length $N$, these last $C$ samples are appended to the beginning of the sequence. This yields a new sequence $x_A^{OFDM}[n]$, $-C \le n \le N-1$, of length $N + C$, where $x_A^{OFDM}[-C], \ldots , x_A^{OFDM}[N-1] = x_A[N-C], \ldots , x_A[N-1], x_A[0], \ldots, x_A[N-1]$. Note that with this definition, $x_A^{OFDM}[n] = x_A[n]_N$ for $-C \le n \le N-1$, which implies that $x_A^{OFDM}[n-k] = x_A[n-k]_N $ for $ -C \le n-k \le N-1$.

Suppose $x_A^{OFDM}[n]$ and $x_B^{OFDM}[n]$ are inputs to a discrete-time channel with impulse response $h_A[n]$ and $h_B[n]$, respectively. The channel output $y_R[n]$, $0 \le n \le N-1$ is then (assuming that the delay spread of FPNC $\max \left[D_A ,D_B \right]$ is no large than the CP length $C$)
%\begin{align}
%\begin{array}{l}
% y_R [n] = x_A^{OFDM} [n]*h_A [n] + x_B^{OFDM} [n]*h_B [n] + w[n] \\
%  = \sum\limits_{k = 0}^{C-1} {h_A [k]x_A^{OFDM} [n - k]}  + \sum\limits_{k = 0}^{C-1} {h_B [k]x_B^{OFDM} [n - k]}  + w[n] \\
%  = \sum\limits_{k = 0}^{C-1} {h_A [k]x_A [n - k]_N }  + \sum\limits_{k = 0}^{C-1} {h_B [k]x_B [n - k]_N }  + w[n] \\
%  = x_A [n] \otimes _N h_A [n] + x_B [n] \otimes _N h_B [n] + w[n],
% \end{array}
%\label{equ:CircularConv1}
%\end{align}
\begin{align}
 y_R [n] &= x_A^{OFDM} [n]*h_A [n] + x_B^{OFDM} [n]*h_B [n] + w[n] \nonumber\\
  &= \sum\limits_{k = 0}^{C-1} {h_A [k]x_A^{OFDM} [n - k]}  + \sum\limits_{k = 0}^{C-1} {h_B [k]x_B^{OFDM} [n - k]}  + w[n] \nonumber\\
  &= \sum\limits_{k = 0}^{C-1} {h_A [k]x_A [n - k]_N }  + \sum\limits_{k = 0}^{C-1} {h_B [k]x_B [n - k]_N }  + w[n] \nonumber\\
  &= x_A [n] \otimes _N h_A [n] + x_B [n] \otimes _N h_B [n] + w[n],
\label{equ:CircularConv1}
\end{align}
where the third equality follows from the fact that for $0 \le k \le C-1$, $x_A^{OFDM}[n-k] = x_A[n-k]_N $ for $ -C \le n-k \le N-1$. Thus, by appending a CP to the channel input, the linear convolution associated with the channel impulse response $y_R[n]$ for $0 \le n \le N-1$ becomes a circular convolution. Taking the DFT of the channel output in the absense of noise then yields the following FPNC frequency domain digital expression:

\vspace{0.1in}
\noindent \textbf{FPNC Frequency Domain Digital Expression:}\vspace{0.1in}\\
Define $C$ as the length of the CP, and assuming FPNC $\text{delay spread} = \max \left[D_A ,D_B \right] \le C$ (where $D_A$ and $D_B$ are functions of the multipath delays $\tau_A^{M_A - 1}$ and $\tau_B^{M_B - 1}$, respectively) the received signal at subcarrier $k$ is given by
%\footnote{We remark here that the subcarriers' indices will be changed from $0, \cdots, N-1$ to $1, \cdots, N$ for the remainder of this paper.}
\begin{align}
Y[k] = H_A [k]X_A [k] + H_B [k]X_B [k] + W[k], \ \ \ \ \ \ k = 0,\ldots,N-1.
\label{equ:FreqModel}
\end{align}
Note that the time-domain delay spread has been incorporated into $H_A[k]$ and $H_B[k]$ respectively. In FPNC, we will map $Y[k]$ for each subcarrier $k$ into the XOR, $X_A [k] \oplus X_B [k]$. This will be detailed in Section 4.3. The main point here is in (\ref{equ:FreqModel}), the signals of different subcarriers $k$ are isolated from each other, and we only need to perform PNC mapping within each subcarrier.

We remark that our discussion so far in this section has assumed the absence of CFO. When there is CFO, inter-carrier interference (ICI) may occur, and this will be further discussed in Section 4.1.

\section{FPNC Frame Format}

\begin{figure}[h]
\centering
\includegraphics[width=1\textwidth]{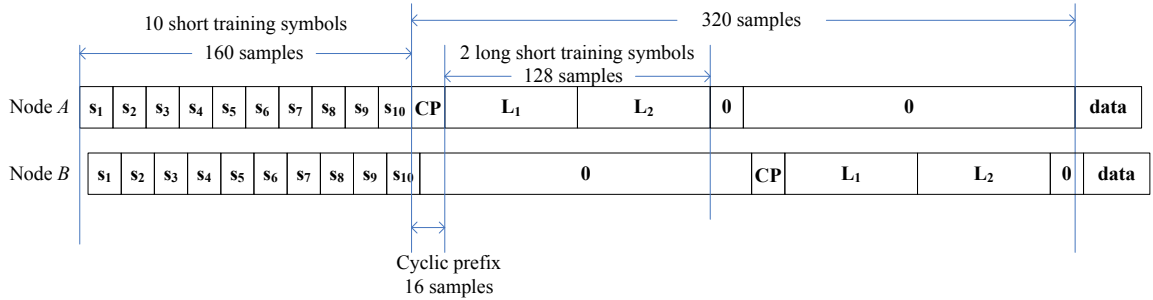}
\caption{FPNC preamble format.}\label{fig:FrameFPNC}
\end{figure}

This section focuses on the PHY frame design to enable asynchronous operation, channel estimation, and frequency offset compensation in FPNC. As previously mentioned, the asynchronous operation requires the PNC delay spread to be within CP. To ensure this, a simple MAC protocol as follows could be used to trigger near-simultaneous transmissions by the two end nodes. The relay could send a short polling frame (similar to the ``beacon frame'' in 802.11 that contains only 10 Bytes) to the end nodes. Upon receiving the polling frame, the end nodes then transmit. With this method, the symbols would arrive at the relay with a relative delay offset of $\left|RTT_A - RTT_B\right|$, where $RTT$ is the round trip time, including the propagation delay and the processing time at the end nodes. This delay offset is not harmful to our system as long as the sample misalignment of two end nodes is within the CP length.

Given this loose synchronization, our training symbols and pilot designs described below can then be used to facilitate channel estimation and frequency offset compensation in FPNC. We modify the PHY preamble design of 802.11a/g for FPNC. The overall FPNC frame format is shown in Fig. \ref{fig:FrameFPNC}. The functions of the different components in the PHY preamble are described in the next few subsections.

%\begin{figure}[h]
%\centering
%\includegraphics[width=0.7\textwidth]{STS.eps}
%\caption{Short Training Symbol design for FPNC (time domain).}\label{fig:STS}
%\end{figure}

\begin{figure*}
   \begin{minipage}[h]{0.5\linewidth}
     \centering
      \includegraphics[width=1\textwidth]{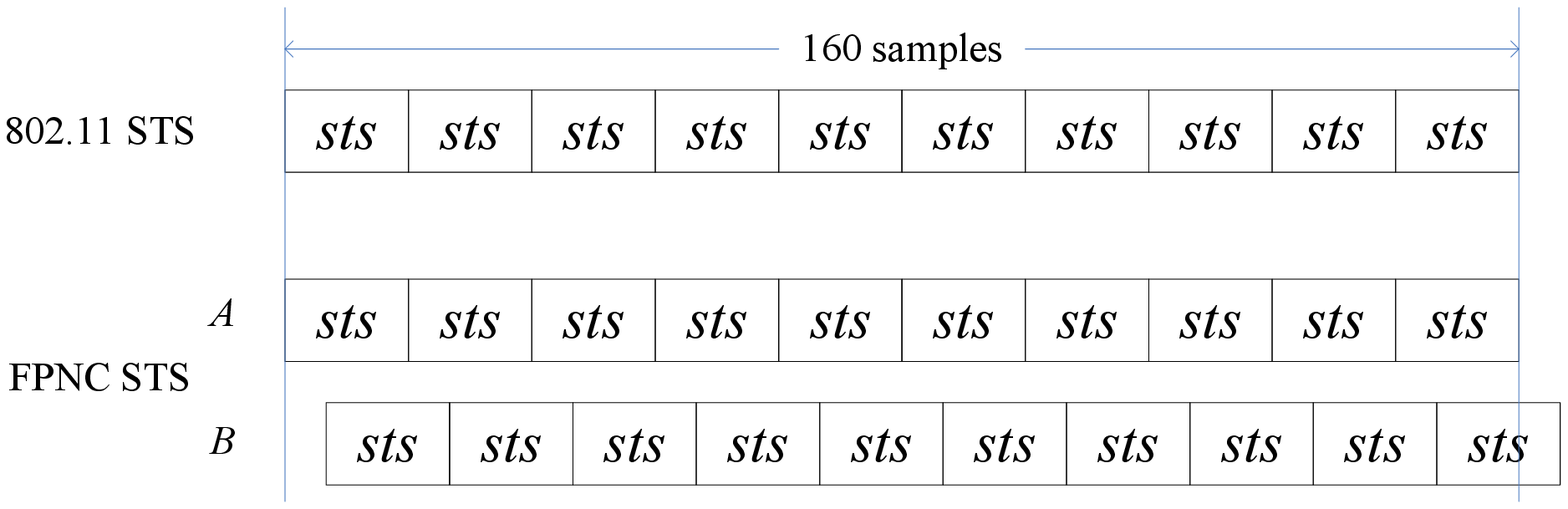}
      \caption{Short Training Symbol design for FPNC (time domain).}
      \label{fig:STS}
   \end{minipage}
   \begin{minipage}[h]{0.5\linewidth}
      \centering
      \includegraphics[width=1\textwidth]{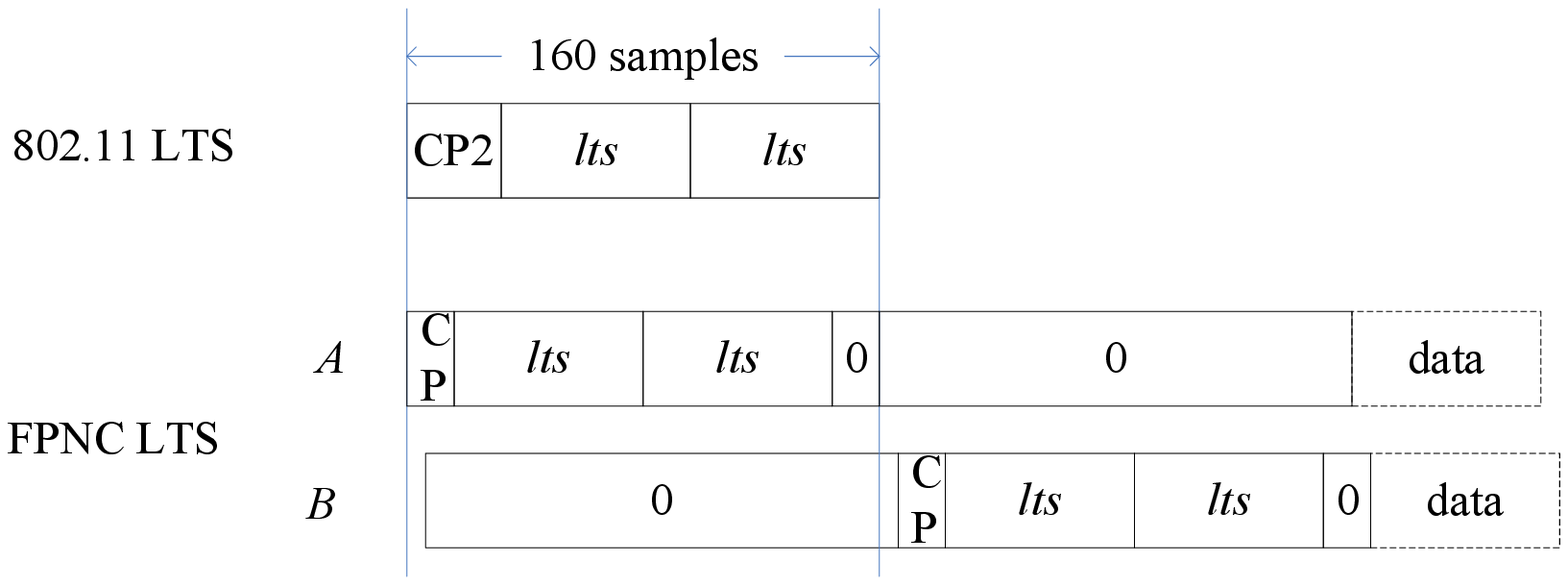}
      \caption{Long Training Symbol design for FPNC (time domain).}
      \label{fig:LTS}
   \end{minipage}
\end{figure*}

\subsection{FPNC Short Training Symbol}
\begin{figure}[h]
\centering
\includegraphics[width=0.7\textwidth]{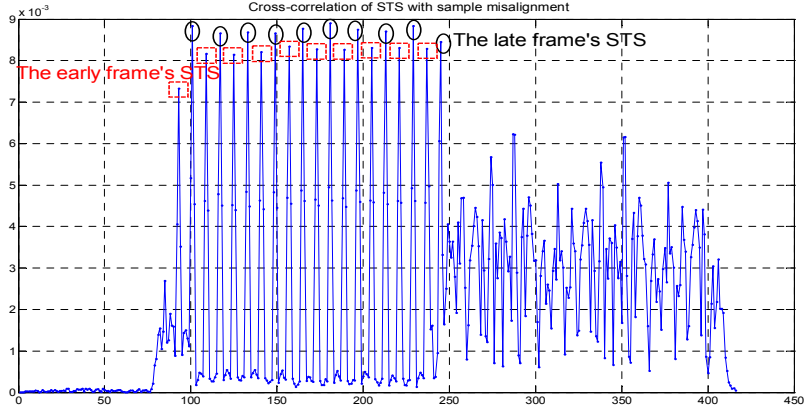}
\caption{Cross-correlation of the STS for the uplink of FPNC.  }\label{fig:CrossCorrelation}
\end{figure}

In 802.11, the short training symbol (STS) sequence contains 160 time-domain samples, in which 16 samples form one STS unit ($sts$) for a total of 10 identical units, as shown in Fig. \ref{fig:STS}. FPNC adopts the same STS sequence as in 802.11, as illustrated in Fig. \ref{fig:STS}. The STS sequence is used by the relay node to perform the sample timing recovery on the received frame. In particular, the relay node applies a cross-correlation to locate the sample boundary for the long training symbols that follow the STS sequence. The normalized cross-correlation is defined as follows:
\begin{align}
Z[n] = \frac{{\left| {\sum\limits_{i = 0}^{L-1} {\left( {sts^* [i]y_R [n + i]} \right)} } \right|}}{{\sum\limits_{i = 0}^{L-1} {\left( {y_R [n + i]y_R^* [n + i]} \right)} }},
\label{equ:CrossCorr}
\end{align}
where $n$ is the received sample index, $y_R[n]$ is the $n$-th sample at the relay $R$, and $L=16$ is the length of each $sts$. For FPNC, this cross-correlation will result in 20 peaks over the STS sequences (see Fig. \ref{fig:CrossCorrelation}) of the two frames if the frames are not synchronized. From this profile of peaks, we can identify the last two peaks. If the Delay-Spread-Within-CP requirement is satisfied, then the last two peaks must be the last peaks of $A$ and $B$, respectively. This is because the CP as well as the sts are of 16 samples in length. From there, we could locate the boundaries of the LTS of $A$ and $B$ that follow. Note that when the STS sequences of nodes $A$ and $B$ overlap exactly, we will have ten peaks only. In this case, the LTS boundaries of $A$ and $B$ also overlap exactly, and we simply use the last peak to identify the common boundary.

%\begin{figure}[h]
%\centering
%\includegraphics[width=0.7\textwidth]{LTS.eps}
%\caption{Long Training Symbol design for FPNC (time domain)}\label{fig:LTS}
%\end{figure}

\subsection{FPNC Long Training Symbol}
With reference to Fig. \ref{fig:LTS}, the 802.11 LTS sequence contains 160 time-domain samples in which there is a CP followed by two identical LTS units, $lts$. The receiver uses the LTS sequence to perform channel estimation and CFO compensation.

For FPNC, in order to estimate two uplink channel gains, we design the LTS so that it contains twice the length of LTS in 802.11a/g, as shown in Fig. \ref{fig:LTS}. In Fig. \ref{fig:LTS}, we intentionally show the case in which the LTS sequences of the two end nodes are not exactly synchronized. Note that we change the 802.11 LTS design by shortening its original CP length from 32 to 16 to make sure that the two $lts$ units of $B$ will not overlap with the data of $A$ that follows under the condition that the delay spread is less than the CP length of 16. This does not impose additional requirement on the delay spread, since the CP of the data OFDM symbols in 802.11a/g (and FPNC) have only 16 samples anyway (i.e., the delay spread must be within 16 samples anyway). Section 4 will detail the CFO compensation and channel estimation methods for our implementation.

\begin{figure}[h]
\centering
\includegraphics[width=0.7\textwidth]{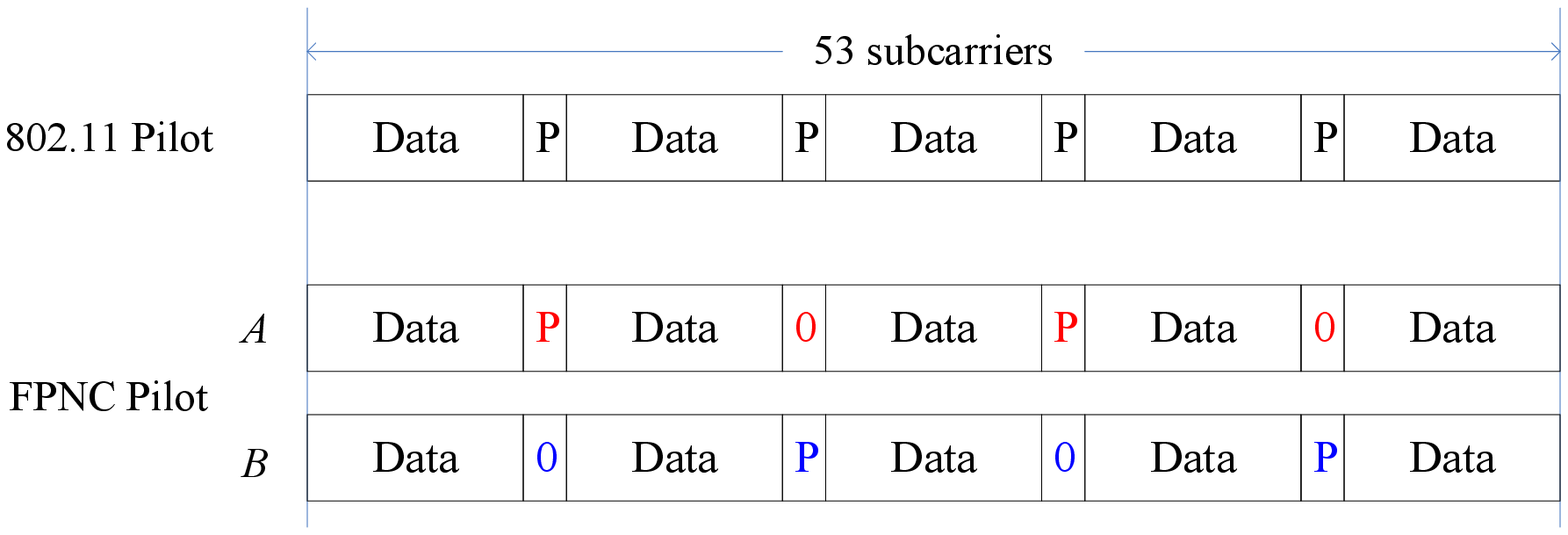}
\caption{Long Training Symbol design for FPNC (time domain).}\label{fig:Pilot}
\end{figure}
\subsection{FPNC Pilot}
There are four pilots for each OFDM symbol in 802.11, as shown in Fig. \ref{fig:Pilot}. The four pilots are used to fine-tune the channel gains estimated from LTS. In a frame, there are multiple OFDM symbols, but only one LTS in the beginning. In practice, the channel condition may have changed by the time the later OFDM symbols arrive at the receiver. That is, the original channel gains as estimated by LTS may not be accurate anymore for the later OFDM symbols. The pilots are used to track such channel changes.

In FPNC, we design the FPNC pilots of nodes $A$ and $B$ by nulling certain pilots to introduce orthogonality between them, as shown in Fig. \ref{fig:Pilot}. As will be detailed in Section 4.2, this allows us to track the channel gains of $A$ and $B$ separately in a disjoint manner in FPNC. We conducted some experiments for a point-to-point communication system using the two-pilot design rather than the four-pilot design. We find that for our linear interpolation channel tracking scheme described in Section 4.2, the BER performances of the two-pilot and four-pilot designs are comparable for BPSK- and QPSK-modulated systems.

\section{Addressing Key Implementation Challenges in FPNC}
We next present our methods for carrier frequency offset compensation, channel estimation, and FPNC mapping, assuming the use of the PHY frame format presented in Section 3.

\subsection{FPNC Carrier Frequency Offset (CFO) Compensation}
For CFO compensation, we first estimate the two independent CFOs (namely $CFO_A$ and $CFO_B$) caused by the carrier frequency offsets between nodes $A$ and $B$ and relay $R$, respectively. We then compensate for the mean of the two CFOs (i.e., $CFO_{PNC}  = \left( {CFO_A  + CFO_B } \right)/2$
). The details are presented below.

\subsubsection{CFO Estimation}
For the uplink phase, when there are CFOs, the received frames at relay R will suffer from time-varying phase asynchronies. We need to compensate for the CFOs to alleviate inter-carrier interference (ICI) among data on different subcarriers.

Recall that in Section 3, we mentioned that a loose synchronization MAC protocol can be used to ensure that the difference of the arrival times of the frames from nodes $A$ and $B$ are within CP.  That means that the LTSs from nodes $A$ and $B$ will overlap with each other substantially, with the non-overlapping part smaller than CP (see Fig. \ref{fig:FrameFPNC}). Recall also that we introduce orthogonality between the LTSs of nodes $A$ and $B$ so that when the LTS units in $A$ are active, the LTS in $B$ are zeros, and vice versa, as shown in Fig. \ref{fig:LTS}. This allows us to separately estimate $CFO_A$ and $CFO_B$. Without loss of generality, in the following we focus the estimation of $CFO_A$ using $LTS_A$.

$CFO_A$ is given by $\Delta f_A = f_A - f_R$ (i.e., the difference in the frequencies of the oscillators of node $A$ and relay $R$). We define the normalized $CFO_A$ to be $\phi _A  = 2\pi \Delta f_A \frac{T}{N}$ , where $T$ is the duration of one OFDM symbol, and $N$ is the number of samples in one OFDM symbol not including CP. In other words, $\phi_A$ is the additional phase advance introduced by the CFO from one sample to the next sample.

To estimate $\phi_A$, we multiply one sample in the first unit of $LTS_A$ (see Fig. \ref{fig:LTS}) by the corresponding sample in the second unit of $LTS_A$ to obtain $\left( {y_R^{LTS_A } [n]} \right)^* y_R^{LTS_A } [n + N]$. Then, $angle\left( {\left( {y_R^{LTS_A } [n]} \right)^* y_R^{LTS_A } [n + N]} \right) \in \left( { - \pi ,\pi } \right)$ is given by
\begin{align}
angle\left( {\left( {y_R^{LTS_A } [n]} \right)^* y_R^{LTS_A } [n + N]} \right) + 2m\pi  = N\phi_A ,
\label{equ:AngleLTS}
\end{align}
where $m \in \{ \ldots, - 2, - 1,0,1,2,\ldots\}$.

For our experimental test-bed, USRP, we found that the accuracies of the onboard oscillators are such that they do not induce large CFOs so that $m=0$ (interested readers are referred to \referred{Savo4Gbook} \cite{Savo4Gbook} for CFO estimation when $m \ne 0$.). Hence, we could write (\ref{equ:AngleLTS}) as follows:
\begin{align}
N\phi _A  = angle\left( {\left( {y_R^{LTS_A } [n]} \right)^* y_R^{LTS_A } [n + N]} \right).
\label{equ:AngleLTS1}
\end{align}
Strictly speaking, (\ref{equ:AngleLTS1}) is an expression for the noiseless case. Because of noise, $angle\left( {\left( {y_R^{LTS_A } [n]} \right)^* y_R^{LTS_A } [n + N]} \right)$ for different $n \in \{ 0, \ldots ,N - 1\} $ could be different. Thus, in our computation, we first obtained $\hat \phi _A [n] = angle\left( {\left( {y_R^{LTS_A } [n]} \right)^* y_R^{LTS_A } [n + N]} \right)$ for $n = 0, \ldots ,N-1$, and then estimate $\phi_A$ by
\begin{align}
\hat \phi _A  = \mathop {median}\limits_{n \in \{ 0, \ldots, N-1\} } \left( {\hat \phi _A [n]} \right).
\label{equ:AngleLTS2}
\end{align}
We obtain $\phi_B$ similarly.

%\com{Do we need????? Fig. CFO. Histograms of CFO for the uplink channel between node A and relay R in one FPNC frame with different SNRs: (a) SNR=10dB; (b) SNR=15dB;(c) SNR=20dB. This plot shows that the mean and median values in the histograms are different, and we should choose median, due to the existence of extreme points that are induced by unknown errors, shown in dotted-red circles.}

\begin{figure}[t]
\centering
\includegraphics[width=1\textwidth]{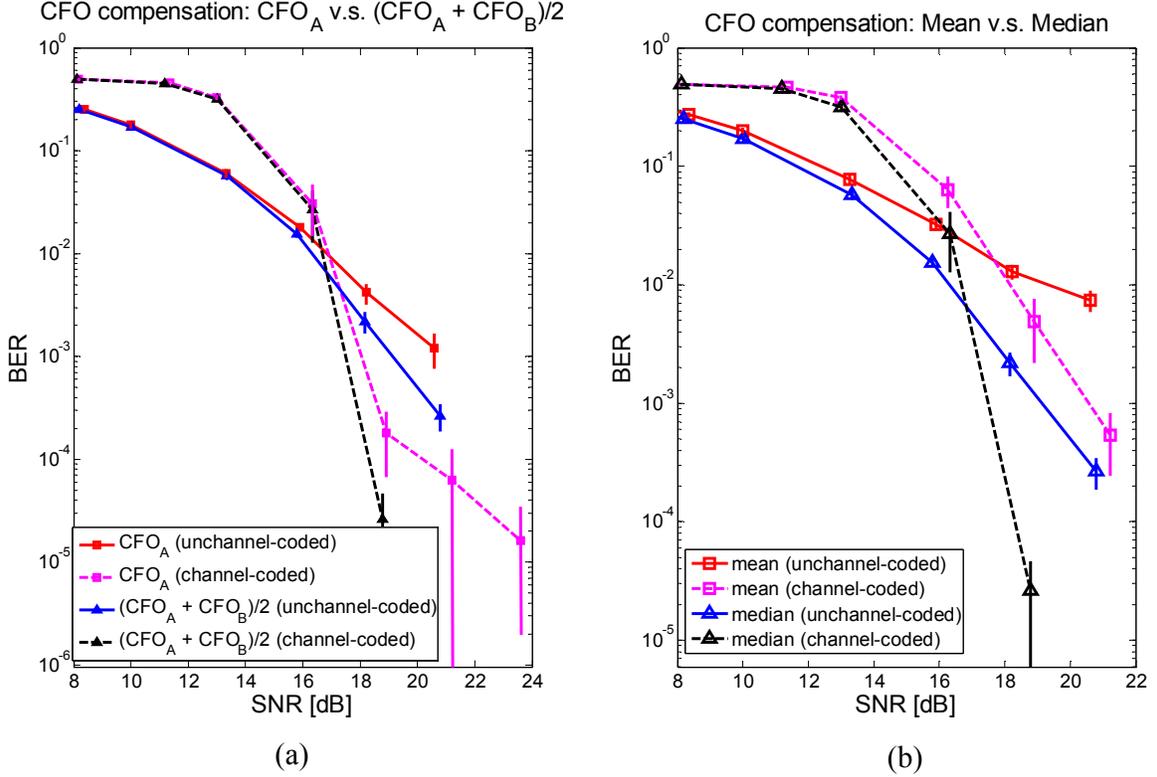}
\caption{Comparison of the different CFO compensation methods: (a) $CFO_A$ v.s. $\frac{CFO_A + CFO_B}{2}$; (b) mean CFO v.s. median CFO.}
\label{fig:CFOMeanMedian}
\end{figure}

The reason we use the median CFO instead of the mean values is that we find the median is more stable. In particular, some samples of $\hat \phi _A [n]$ are outliers that appear to be caused by unknown errors of significant magnitudes. We will show the BER results comparing the use of mean and median for CFO compensation (in Fig. \ref{fig:CFOMeanMedian}(b)).
%Fig. \ref{fig:CFOMeanMedian}(a) presents the performance difference between the two methods.

\subsubsection{Compensation for Two CFOs}

%Fig. \cite{CFO}\textcolor{red}{Put the plot instead of the Table} shows $CFO_A$ and $CFO_B$. If left uncompensated, the received sample will suffer a ICI on the order of 10 kHz (e.g., $\Delta f = \frac{{\phi _A [n]}}{{2\pi t}}$, where $t = 50$ns), which will diminish the performance of coherent detection since the gap between each subcarrier is 375 kHz for 802.11.
%\footnote{In fact, one could apply more sophisticated methods to cope with the two ICIs due to imperfect CFO compensation. For instance, one may want to use ZF (zero forcing) or even a MAP (maximum a posterior) decoder, which inevitably will introduce some processing cost. However, detailed studies and performance comparisons among different ICI compensation methods are beyond the scope of this paper, which are left as a future work.\com{Do we still want this footnote?}}

In FPNC, we adopt the mean of the two CFOs for compensation purposes:
\begin{align}
\tilde \phi  = \left( {\hat \phi _A  + \hat \phi _B } \right)/2.
\label{equ:CFO}
\end{align}
Experimental results show (see Fig. \ref{fig:CFOMeanMedian}(a)) that compensation by the mean $\tilde \phi$ in (\ref{equ:CFO}) is better than compensation by either $\hat \phi _A$ or $\hat \phi _B$. We believe a theoretical study to explore and compare different compensatation methods may be worthwhile in the future. As far as we know, there  have been no theoretical treatments of compensating for two CFOs. Fig. \ref{fig:CFOMeanMedian}(b) shows the BER performance of using median for the estimate of $\hat \phi _A$ or $\hat \phi _B$ as in (\ref{equ:AngleLTS2}), versus using mean. It shows that he use of median results in better performance.

After compensation, our received data in the time domain is given by
\begin{align}
\tilde y_R [n] = y_R [n]e^{ - jn\tilde \phi }.
\label{equ:CFO1}
\end{align}
In the frequency domain, we have
\begin{align}
\tilde Y_R [k] = DFT(\tilde y_R [n]).
\label{equ:CFO2}
\end{align}
We should emphasize that the computation complexity of FPNC CFO compensation is exactly the same as that of point-to-point communication, thus real-time decoding is possible.

\subsection{FPNC Channel Estimation}
In this subsection, we present the channel estimation and tracking method for FPNC. Note that CFO compensation was performed on the time-domain signal. For channel estimation, however, we are interested in the channel gains for different subcarriers in the frequency domain. This means that channel estimation will be performed after DFT. Thus, in the following we look at the signal after CP removal and DFT.

For FPNC channel estimation, we use the LTS to obtain a first estimate. Pilots are used to obtain additional estimates for later OFDM symbols within the same frame. In the following, we consider channel estimation of $H_A[k]$. Estimation of $H_B[k]$ is performed similarly.

For channel estimation based on LTS, define one FPNC LTS unit of node $A$ (i.e., with respect to Fig. \ref{fig:LTS}, one unit is $lts_A$) in the frequency domain as $X_A^{LTS} [k]$, where $k = 0, \ldots, N-1$. Based on the first unit of $lts_A$ the received frequency domain $LTS_A$ (i.e., $\tilde Y_R^{LTS_A } [k] = DFT\left( {\tilde y_R^{LTS_A } [n]} \right)$),  we perform channel estimation of $H_A[k]$ as follows:
\begin{align}
\hat H_A [k] = \frac{{\tilde Y_R^{LTS_A } [k]}}{{X_A^{LTS} [k]}}.
\label{equ:ChannelEst}
\end{align}

As mentioned in Section 3, each LTS contains two identical units in our design. The uplink channel gain $H_A[k]$ between node $A$ and relay $R$ is estimated by taking the average of the two units results
\begin{align}
\tilde H_A [k] = \left( {\hat H_A [k] + \hat H_A [k + N]} \right)/2.
\label{equ:ChannelEst1}
\end{align}

In general, the channel may have changed from the first OFDM symbol to the last OFDM symbol within the same frame. The estimate based on LTS in (\ref{equ:ChannelEst1}) applies only for the earlier symbols. Pilots are used to track the channel changes for later symbols. Our pilot design was shown in Fig. Pilot. In each FPNC OFDM symbol, there are two pilots per end node. Note from Fig. \ref{fig:Pilot} that the two pilots of node $A$ and the two pilots are node $B$ are positioned at different subcarriers and non-overlapping in the frequency domain. Therefore, we could separately track the changes in $H_A[k]$ and $H_B[k]$. In the following, we consider the tracking of $H_A[k]$. Tracking of $H_B[k]$ can be done similarly.

Let $k'$ and $k''$ denote the subcarriers occupied by the two pilots of $A$. Consider OFDM symbol $m$. Let $\tilde Y_R^m [k']$ and $\tilde Y_R^m [k'']$ be the received signal in the frequency domain. Because the pilots of $A$ and $B$ do not overlap, $\tilde Y_R^m [k']$ and $\tilde Y_R^m [k'']$ contain only signals related to the pilots of $A$. We first multiply $\tilde Y_R^m [k']$ and $\tilde Y_R^m [k'']$ by $(\tilde H_A^m [k'])^{ - 1}$ and $(\tilde H_A^m [k''])^{ - 1}$ obtained from (\ref{equ:ChannelEst1}), respectively. Let $P_A [k']$ and $P_A [k'']$ be the two pilots. Then, we compute
\begin{align}
\begin{array}{l}
 \Delta \tilde H_A^m [k'] = (\tilde H_A^m [k'])^{ - 1} {{\tilde Y_R^m [k']} \mathord{\left/
 {\vphantom {{\tilde Y_R^m [k']} {P_A [k']}}} \right.
 \kern-\nulldelimiterspace} {P_A [k']}}, \\
 \Delta \tilde H_A^m [k''] = (\tilde H_A^m [k''])^{ - 1} {{\tilde Y_R^m [k'']} \mathord{\left/
 {\vphantom {{\tilde Y_R^m [k'']} {P_A [k'']}}} \right.
 \kern-\nulldelimiterspace} {P_A [k'']}}.
\end{array}
\label{equ:ChannelEst2}
\end{align}

After that, we perform linear fitting to obtain $\Delta \tilde H_A^m [k]$ for $k \ne k',k''$ , as follows:
\begin{align}
\Delta \tilde H_A^m [k] = \Delta \tilde H_A^m [k'] + \left( {\frac{{\Delta \tilde H_A^m [k''] - \Delta \tilde H_A^m [k']}}{{k'' - k'}}} \right)(k - k').
\label{equ:ChannelEst3}
\end{align}

To obtain the final channel estimation for the $m$-th OFDM symbol, we compute
\begin{align}
H_A^m [k] = \tilde H_A^m [k] \cdot \Delta \tilde H_A^m [k].
\label{equ:ChannelEst4}
\end{align}

\begin{figure}[h]
\centering
\includegraphics[width=0.7\textwidth]{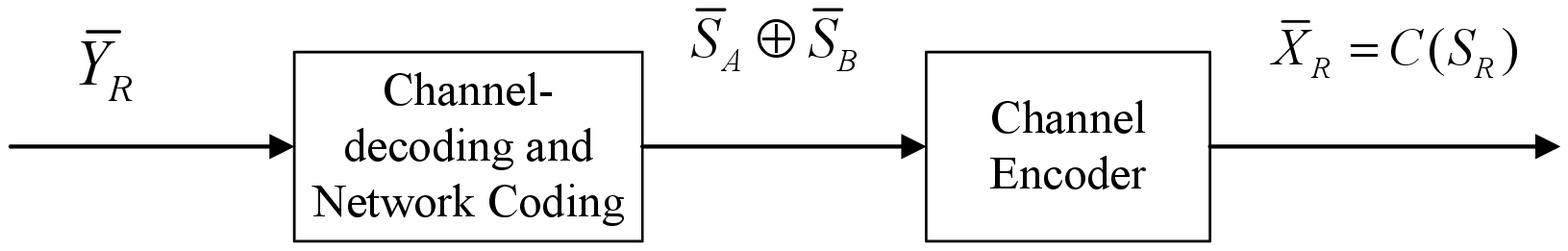}
\caption{Link-by-link channel-coded PNC, including channel-decoding and network coding (CNC) process and channel encoding.}\label{fig:LinkbyLinkPNC}
\end{figure}

\subsection{FPNC Mapping}
For reliable communication, channel coding should be used. Channel coding in PNC systems can be either done on an end-to-end basis or a link-by-link basis \referred{ShengliJSAC09,PNCSurvey}\cite{ShengliJSAC09,PNCSurvey}. The latter generally has better performance because the relay performs channel decoding to remove noise before forwarding the network-coded signal.

The basic idea in link-by-link channel-coded PNC is shown in Fig. \ref{fig:LinkbyLinkPNC}. It consists of two parts. Let $\bar Y_R$ denote the vector representing the overall channel-coded overlapped frames received by relay $R$. The operation performed by the first part is referred to as the Channel-decoding and Network-Coding (CNC) process in \referred{ShengliJSAC09}\cite{ShengliJSAC09}. It maps $\bar Y_R$ to $\bar S_A \oplus \bar S_B$, where $\bar S_A$ and $\bar S_B$ are the vectors of source symbols from nodes $A$ and $B$, respectively, and the $\oplus$ operation represents symbol-by-symbol XOR operation across corresponding symbols in $\bar S_A$ and $\bar S_B$. Note that the number of symbols in $\bar Y_R$ is more than the number of symbols in $\bar S_A \oplus \bar S_B$ because of channel coding. Importantly, CNC involves both channel decoding and network coding. In particular, CNC channel-decodes the received signal $\bar Y_R$ not to $\bar S_A$ and $\bar S_B$ individually, but to their XOR.  The second part can be just any conventional channel coding operation that channel code $\bar S_A  \oplus \bar S_B $ to $\bar X_R  = C(\bar S_A  \oplus \bar S_B )$ for broadcast to nodes $A$ and $B$, where $C(*)$ is the channel coding operation.

As mentioned in \referred{ShengliJSAC09}\cite{ShengliJSAC09} and \referred{PNCSurvey}\cite{PNCSurvey}, the CNC component is unique to the PNC system, and different designs can have different performance and different implementation complexity. We refer the interested readers to \referred{PNCSurvey}\cite{PNCSurvey} for a discussion on different CNC designs.

\begin{figure}[h]
\centering
\includegraphics[width=0.7\textwidth]{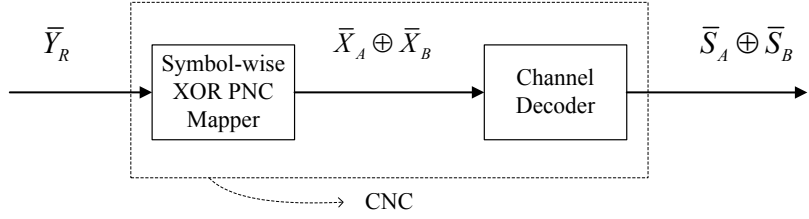}
\caption{XOR-CD design for CNC.}\label{fig:XORCDFlowChart}
\end{figure}

In this paper, we choose a design that is amenable to simple implementation, as shown in Fig. \ref{fig:XORCDFlowChart}. We refer to this CNC design as XOR-CD. In this design, any linear channel code can be used. In our implementation, we choose to use the convolutional code. In XOR-CD, the channel-decoding and network coding operations in CNC are performed in a disjoint manner. As shown in Fig. \ref{fig:XORCDFlowChart}, based on the CFO-compensated $\tilde Y_R[k]$ obtained as in (\ref{equ:CFO2}), we obtain the overall vector $\bar Y_R = \left( {Y_R [k]} \right)_{k = 0,1, \ldots}$. We then perform symbol-wise PNC mapping to get an estimate for the the channel-coded XOR vector $\bar X_A \oplus \bar X_B  = \left( {X_A [k] \oplus X_B[k]} \right)_{k = 0,1, \ldots}$, where $\bar X_A  = \left( {X_A [k]} \right)_{k = 0,1, \ldots}$ and $\bar X_B  = \left( {X_B [k]} \right)_{k = 0,1, \ldots}$ are the channel-coded vectors from $A$ and $B$, respectively. We assume the same linear channel code is used at nodes $A$, $B$, and $R$. Note that since we adopt the convolutional code, $C(*)$ is linear. Therefore, we have $\bar X_A  \oplus \bar X_B  = C(\bar S_A ) \oplus C(\bar S_B ) = C(\bar S_A  \oplus \bar S_B )$, and thus the same Viterbi channel decoder as used in a conventional point-to-point communication link can be used in the second block of Fig. \ref{fig:XORCDFlowChart}.

The mapping in the first block in Fig. \ref{fig:XORCDFlowChart} could be performed as follows. Based on the channel gains estimated in (\ref{equ:ChannelEst4}), we could perform the XOR mapping for the $k$-th subcarrier in the $m$-th OFDM symbol (assuming BPSK modulation) according to the decision rule below:
\begin{align}
\begin{array}{l}
 \exp \left\{ { - \frac{{\left| {Y_R^m [k] - H_A^m [k] - H_B^m [k]} \right|^2 }}{{2\sigma ^2 }}} \right\} + \exp \left\{ { - \frac{{\left| {Y_R^m [k] + H_A^m [k] + H_B^m [k]} \right|^2 }}{{2\sigma ^2 }}} \right\} \\
  \mathop \gtrless \limits^{X_R^m [k] =  1} \limits_{X_R^m [k] =  - 1}
 \exp \left\{ { - \frac{{\left| {Y_R^m [k] + H_A^m [k] - H_B^m [k]} \right|^2 }}{{2\sigma ^2 }}} \right\} + \exp \left\{ { - \frac{{\left| {Y_R^m [k] - H_A^m [k] + H_B^m [k]} \right|^2 }}{{2\sigma ^2 }}} \right\},
 \end{array}
 \label{equ:PNCMapping}
\end{align}
where we have assumed Gaussian noise with variance $\sigma ^2$. The computation complexity in (\ref{equ:PNCMapping})\footnote{Note that (\ref{equ:PNCMapping}) is similar to (7) in Ref. \referred{UPNCICC11}\cite{UPNCICC11}, except that here we allow for the possibility that $|H_A^m [k]| \ne |H_B^m [k]|$}, however, is large. In our implementation, we adopt a simple ``log-max approximation'' \referred{BoydCOBook}\cite{BoydCOBook} (i.e., $\log (\sum\nolimits_i {\exp (z_i )}) \approx \mathop {\max }\limits_i z_i $) that yields the following decision rule:
\begin{align}
\begin{array}{l}
 \min \left\{ {\left| {Y_R^m [k] - H_A^m [k] - H_B^m [k]} \right|^2 ,\left| {Y_R^m [k] + H_A^m [k] + H_B^m [k]} \right|^2 } \right\} \\
  \mathop \gtrless \limits^{X_R^m [k] =  -1} \limits_{X_R^m [k] =  1}
 \min \left\{ {\left| {Y_R^m [k] + H_A^m [k] - H_B^k [k]} \right|^2 ,\left| {Y_R^m [k] - H_A^m [k] + H_B^m [k]} \right|^2 } \right\}.
 \end{array}
 \label{equ:PNCMapping1}
\end{align}
This decision rule can also be interpreted as in Table \ref{tab:PNCmapping}, where
\begin{align}
U \buildrel \Delta \over = \mathop {\arg }\limits_{U \in \{  \pm H_A^m [k] \pm H_B^m [k]\} } \min \left\{ {\left| {Y_R^m [k] - U} \right|^2 } \right\}.
\label{equ:PNCMapping2}
\end{align}

\begin{table}[h]
\caption{XOR mapping with BPSK modulation in FPNC.}
\centering
\begin{tabular}{|c|c|c|}
\hline \multicolumn{1}{|c|}{$U = \mathop {\arg }\limits_{U \in \{  \pm H_A^m [k] \pm H_B^m [k]\} } \min \left\{ {\left| {Y_R^m [k] - U} \right|^2 } \right\}$} &\multicolumn{1}{|c|}{} &\multicolumn{1}{|c|}{$X_R^m [k] = X_A^m [k] \oplus X_B^m [k]$}
\\\hline $H_A^m [k] + H_B^m [k]$ & & 1
\\\hline $H_A^m [k] - H_B^m [k]$ & & -1
\\\hline $-H_A^m [k] + H_B^m [k]$ & & -1
\\\hline $-H_A^m [k] - H_B^m [k]$ & & 1
\\\hline
\end{tabular}
\tabcolsep -10mm\label{tab:PNCmapping}
\end{table}

Note here that this decision rule could be used even for  non-Gaussian noise. This is because (\ref{equ:PNCMapping2}) corresponds to finding the  nearest point in the constellation map (constructed by combining the two end nodes' channel gains).
%Thus, strictly speaking, the assumption of Gaussian noise is not necessary for the decision rule (\ref{equ:PNCMapping2}).

Based on the XORed samples detected using the decision rule of Table \ref{tab:PNCmapping}, we then perform the channel decoding to get the XORed source samples. In our implementation, we use a Viterbi decoder with hard input and hard output. In general, a soft Viterbi algorithm  could also be used for potentially better BER performance \referred{SklarDCbook}\cite{SklarDCbook}.

%The above discussions have addressed the implementation challenges in practicing FPNC, based on which, we now present our implementation methodology and the experimental results for FPNC in the following section.

\section{Experimental Results}
This section presents details of our FPNC implementation over the software radio platform and the experimental results.
\subsection{FPNC Implementation over Software Radio Platform}
We implement FPNC in a 3-node GNU Radio testbed, with Software Defined Radio (SDR). The topology is shown in Fig. \ref{fig:SystemModel}. Each node is a commodity PC connected to a USRP GNU radio \referred{GNURadio}\cite{GNURadio}.
\begin{itemize}
\item \textbf{Hardware:} We use the Universal Software Radio Peripheral (USRP) \referred{Ettus}\cite{Ettus} as our radio hardware. Specifically, we use the XCVR2450 daughterboard operating in the 2.4/5GHz range as our RF frontend. We use the USRP1 motherboard for baseband data processing. The largest bandwidth that USRP1 could support is 8MHz. In our experiment, we use only use half of the total bandwidth for FPNC (i.e., 4MHz bandwidth).

\item \textbf{Software:} The software for baseband signal processing is based on the open source of GNURadio project \referred{GNURadio}\cite{GNURadio}. We build our system by  modifying the 802.11g transmitter implementation in the FTW project \referred{FTW}\cite{FTW}. The FTW project \referred{FTW1}\cite{FTW1}, however, does not have a 802.11g receiver.  Therefore, we develop our own OFDM receiver, designed specifically to tackle various issues in the FPNC system, such as CFO estimation and compensation, channel estimation, and CNC processing as presented in Section 4.

\end{itemize}

\subsection{Experimental Results}
We conduct our experiments over the channel one of 802.11g, with 2.412GHz being the central frequency. For each transmitter power level (we vary the SNR from 5dB to 20dB), we transmit 1000 packets and examine the resulting BER performance. Both the symbol-synchronous and symbol-asynchronous cases are investigated. The packet length is 1500Bytes, which is a normal Ethernet frame size.
\subsubsection{Time-Synchronous FPNC versus Time-Asynchronous FPNC}
In Section 2, we derived theoretically that as long as the Delay-Spread-Within-CP requirement is satisfied, FPNC will not have asynchrony in the frequency domain. Of interest is whether this reduces the asynchrony penalty in practice. In our first set of experiments, we investigate this issue. We study both unchannel-coded as well as channel-coded FPNC systems.

\begin{figure}[h]
\centering
\includegraphics[width=1\textwidth]{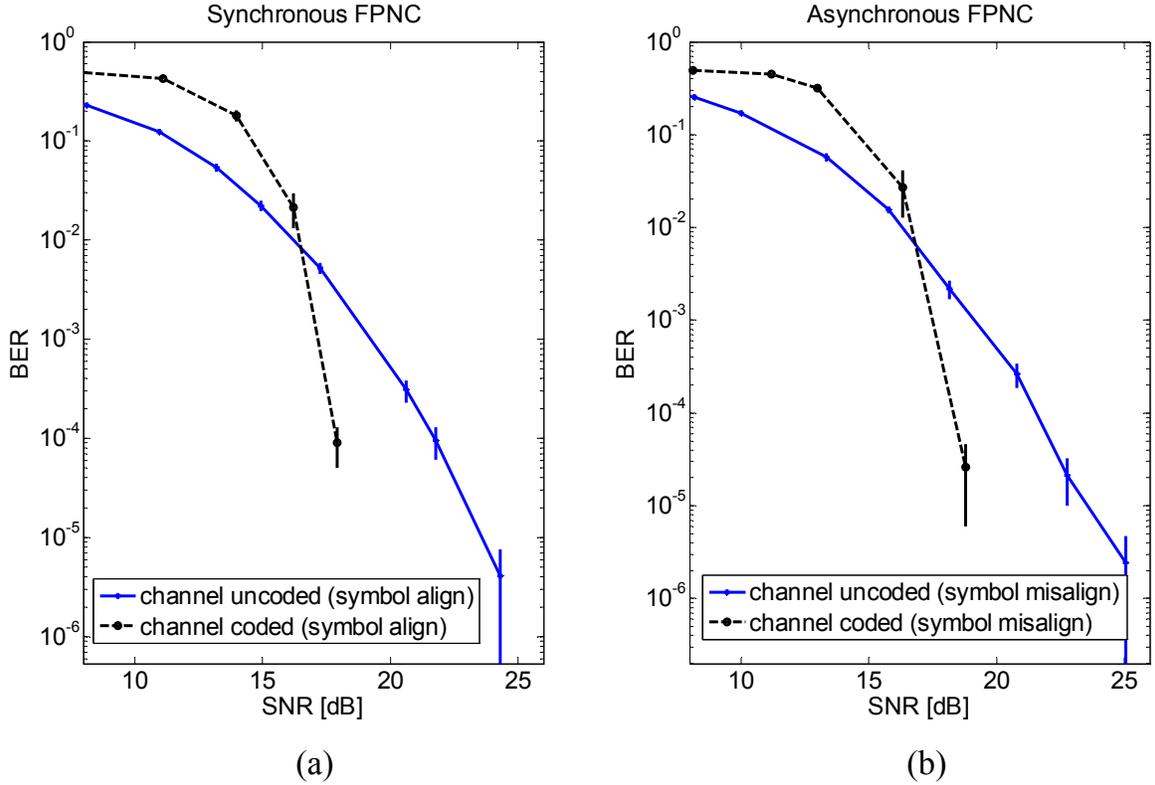}
\caption{BER of FPNC with and without sample synchronization. The 95\% confidence intervals are marked in the figures. Note that the BER here is related to whether the XOR bit is decoded correctly, not whether the individual bits from the two end nodes are decoded correctly.}\label{fig:BerFPNC}
\end{figure}

To create different levels of time asynchrony, we adjust the positions of the end nodes. One of the set-ups corresponds to the perfectly synchronized case (the STS correlation has only ten peaks in the perfectly synchronized case: see Section 3).  Fig. \ref{fig:BerFPNC}(a) shows the BER-SNR curves for the synchronous case, and Fig. \ref{fig:BerFPNC}(b) shows the curves for the asynchronous case with eight samples offset between the early and late frames. Note that this asynchrony still satisfies the Delay-Spread-Within-CP requirement because the CP has of 16 samples. We find that the performance results of the asynchronous cases with other time offset to be similar, and we therefore present the results of the eight-sample offset only.

From Fig. \ref{fig:BerFPNC}(a) and (b), we see that the asynchronous FPNC has essentially the same BER performance as that of the synchronous FPNC. Hence, we conclude that FPNC is robust against time asynchrony as far as BER performance is concerned.

\subsubsection{FPNC versus Other Approaches for TWRC }
Our next set of experiments is geared toward the comparison of FPNC with other TWRC schemes. Recall that FPNC TWRC is a two-phase scheme using two time slots for the exchange of a pair of packets between two end nodes. We consider the following two additional approaches \referred{PNCSurvey}\cite{PNCSurvey}
\begin{itemize}
\item \textbf{SNC:} The straightforward network coding (SNC) scheme makes use of conventional network coding at the higher layer using three time slots. In SNC, node $A$ transmits to relay $R$ in the first time slot; node $B$ transmits to relay $R$ in the second time slot; relay $R$ then XOR the two packets from $A$ and $B$ and transmits the XOR packets to nodes $A$ and $B$ in the third time slot.

\item \textbf{TS:} Traditional scheduling (TS) scheme uses four time slots. In the first time slot, node $A$ transmits to relay $R$; in the second time slot, relay $R$ forwards the packet from $A$ to node $B$. Similarly, the packet from node $B$ to node $A$ uses two additional times slots for its delivery.

\end{itemize}

Our overall goal is the compare the throughputs of the three schemes. To derive the throughputs, we first measure the following three frame-error rates:
\begin{enumerate}[label= \arabic*)]
\item $FER_{PNC}  = P_f^{uplink,PNC} $: frame-error rate of the uplink of FPNC.
\item $FER_{SNC}  = P_f^{uplink,SNC} $: frame-error rate computed from the two uplink time slots in SNC
\item $FER_{P2P}  = P_f^{P2P} $: frame-error rate of a point-to-point link.
\end{enumerate}
%The same convolutional channel code is used in all three cases (see Section 4.3).
Channel coded systems are considered in our implementation. All three systems use the convolutional channel code with 1/2 coding rate, as specified in the 802.11a/g standard \referred{dot11std07}\cite{dot11std07}.

\begin{figure}[tt]
\centering
\includegraphics[width=1\textwidth]{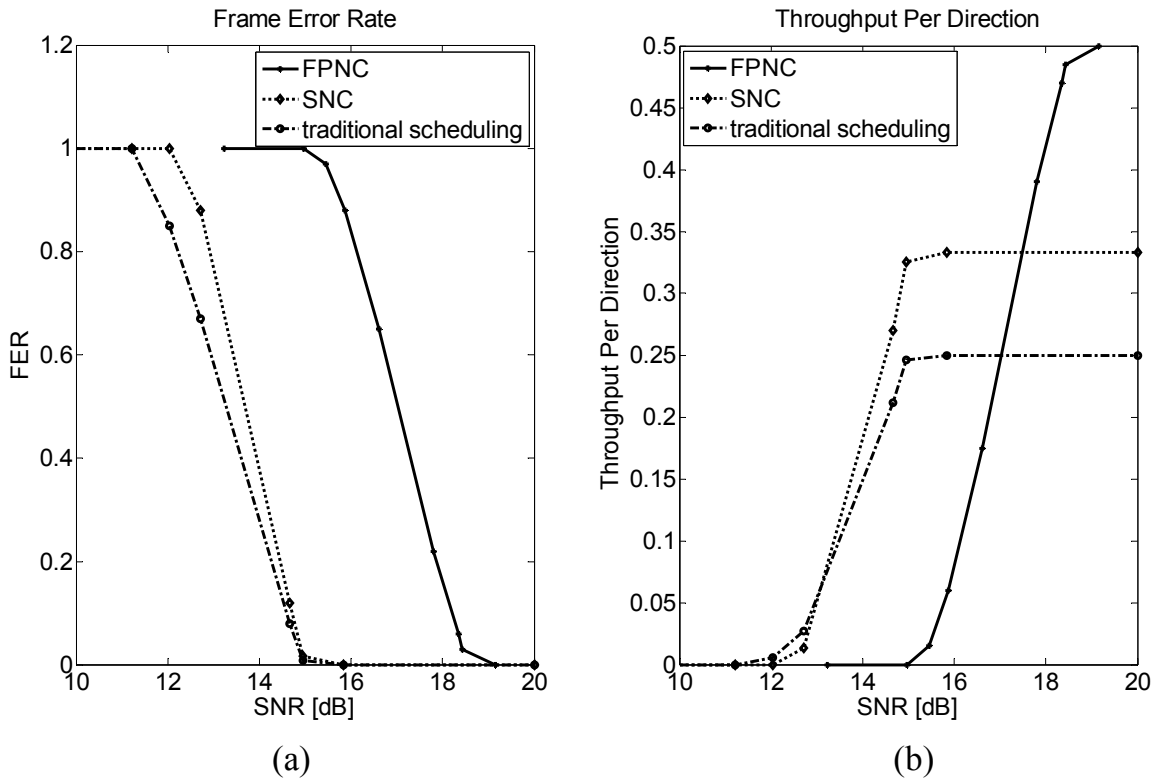}
\caption{Frame error rate and throughput comparison of FPNC with straight-forward network coding and traditional secluding. (a) FER comparison of three approaches; (b) throughput comparison of three approaches.}\label{fig:FerFPNC}
\end{figure}

Note for 1) and 2), $FER_{PNC}$ and $FER_{SNC}$ refers to the error rate of the XOR of the two source frames. That is the error rate for the frame $\bar S_R  = \bar S_A  \oplus \bar S_B $. For $FER_{SNC}$, we gather the decoded $\bar S_A$ and $\bar S_B$ from the two uplink time slots, and the compute their XOR before checking whether there is an error in the XOR frame. For $FER_{PNC}$, the CNC scheme as described in Section 4.3 is used to decode $\bar S_R$ directly based on the simultaneously received signals. In Fig. \ref{fig:FerFPNC}(a), we plot $FER_{PNC}$, $FER_{SNC}$, and $FER_{P2P}$ versus SNR obtained from our experiments. The FER measurements are all from channel-coded systems.
%For completeness, we also plot the corresponding BER for the channel-coded and unchannel-coded systems in Fig. \ref{fig:BerComparison}.

The throughputs per direction of the three TWRC schemes are computed as follows:
%\begin{align}
%\begin{array}{l}
%Th_{FPNC}  = \frac{1}{2}(1 - P_f^{uplink,PNC} )(1 - P_f^{P2P} ) \\
%Th_{SNC}  = \frac{1}{3}(1 - P_f^{uplink,SNC} )(1 - P_f^{P2P} ) \\
%Th_{TS}  = \frac{1}{4}(1 - P_f^{P2P} )^2
%\end{array} \label{equ:Throughput}
%\end{align}
\begin{align}
Th_{FPNC} &= \frac{1}{2}(1 - P_f^{uplink,PNC} )(1 - P_f^{P2P} ), \nonumber \\
Th_{SNC}  &= \frac{1}{3}(1 - P_f^{uplink,SNC} )(1 - P_f^{P2P} ), \nonumber \\
Th_{TS}   &= \frac{1}{4}(1 - P_f^{P2P} )^2.
\label{equ:Throughput}
\end{align}

%\begin{figure}[h]
%\centering
%\includegraphics[width=0.6\textwidth]{BerComparison.eps}
%\caption{BER comparison of FPNC, SNC and TS with and without channel codes.}\label{fig:BerComparison}
%\end{figure}

In Fig. \ref{fig:FerFPNC}(b), we plot the throughputs ($Th_{FPNC}$, $Th_{CNC}$, and $Th_{TS}$) of FPNC, SNC and TS versus SNR based on the $FER_{FPNC}$, $FER_{CNC}$, and $FER_{TS}$ in Fig. \ref{fig:FerFPNC}(a). With reference to Fig. \ref{fig:FerFPNC}(b), for the high SNR regime (above 19dB), the throughput of PNC is approximately 99\% higher than that of the TS scheme, and 49\% higher than that of the SNC scheme. This is essentially the same as the ideal 100\% and 50\% throughput gains derived by slot counting in \referred{PNC06}\cite{PNC06} (i.e., the error-free case), with the difference that we have channel coding here to ensure reliable communication. If we use the guideline that the common decodable 802.11 link usually works at an SNR regime that is higher than 20dB \referred{GeierSNR, KattiANC07}\cite{GeierSNR, KattiANC07}, we can conclude that our FPNC implementation has very good performance in this regime. We note that for this regime, \referred{GeierSNR}\cite{KattiANC07} mentions that ANC can achieve 70\% and 30\% throughput gains relative to TS and SNC. Hence, FPNC has better performance in this SNR regime by comparison.

We note from Fig. \ref{fig:FerFPNC}(b) that the performance of FPNC is not as good as that of SNC or TS at the lower SNR regime (say below 17dB).  This is most likely due to our specific implementation of FPNC in this paper rather than a fundamental limitation of FPNC in general. In particular, recall that we implement the CNC function in FPNC mapping (see Section 4.3) using the so-called XOR-CD approach. In XOR-CD, (i) we first perform XOR mapping for the channel-coded symbol pairs from the two end nodes; (ii) after that channel decoding is applied on the channel-coded XOR symbols to get the XOR of the source symbols. Step (i) loses information that could be useful for the decoding of the XOR of the source symbols, and may lead to inferior performance in the low SNR regime. This phenomenon is explained in \referred{ShengliJSAC09, PNCSurvey }\cite{ShengliJSAC09, PNCSurvey}, and an joint CNC scheme \referred{APNC, PNCSurvey}\cite{APNC, PNCSurvey} for the PNC system can potentially achieve better performance than the XOR-CD scheme implemented in this paper, at the cost of higher implementation complexity.

%\begin{figure}[tt]
%\centering
%\includegraphics[width=1\textwidth]{FerFPNC.eps}
%\caption{Frame error rate and throughput comparison of FPNC with straight-forward network coding and traditional secluding. (a) FER comparison of three approaches; (b) throughput comparison of three approaches.}\label{fig:FerFPNC}
%\end{figure}

\section{Conclusion and Future Work}

This paper presents the first implementation of a PNC system in which the relay performs the XOR mapping on the simultaneously received signals as originally envisioned in \referred{PNC06}\cite{PNC06}. In particular, in our implementation, the XOR mapping is performed in the frequency domain of an OFDM PNC system. We refer to the OFDM PNC system as FPNC. The implementation of FPNC requires us to tackle a number of implementation challenges, including carrier frequency offset (CFO) compensation, channel estimation, and FPNC mapping.

A major advantage of FPNC compared with PNC in the time domain is that FPNC can deal with the different arrival times of the signals from the two end nodes in a natural way. We show by theoretical derivation that if the simultaneously received signals in FPNC have a maximum delay spread that is less than the length of the OFDM cyclic prefix (CP), then after the Discrete Fourier Transform, the frequency-domain signals on the different subcarrier are isolated from each other. That is, in the frequency domain, the signals are synchronous. Then, straightforward XOR mapping can be applied on the different subcarrier signals separately in a disjoint manner. To validate the advantage of FPNC, we present experimental results showing that time-domain symbol asynchrony does not cause performance degradation in FPNC.

To date, most work related to PNC focuses on its potential superior performance as derived from theory. In this paper, we evaluate the throughput gain of PNC relative to other two-way relay schemes. Our implementation indicates that PNC can have a throughput gain of 99\% compared with traditional scheduling (TS), and a 49\% throughput gain compared with strait-forward network coding (SNC), in the high SNR regime (above 19 dB) in which practical technology such as Wi-Fi typically operates.

Going forward, there are many rooms for improvement in our FPNC implementation. In this paper, when faced with alternative design choices, we opt for implementation simplicity than performance superiority. For example, we choose to use a simple PNC mapping method called XOR-CD in this paper, which is simple to implement but has inferior performance compared with other known methods \referred{PNCSurvey}\cite{PNCSurvey} in the low SNR regime. In addition, our implementation exercise reveals a number of problems with no good theoretical solutions yet, and further theoretical analysis is needed; in such cases, we use simple heuristics to tackle the problems. For example, CFO compensation for FPNC is an area that is not well understood yet, because we have to deal with CFOs of more than one transmitter relative to the receiver. In this paper, we simply compensate for the mean of the CFOs of the two end nodes. Better methods await further theoretical studies. Last but not least, we base our design on the 802.11 standard to a large extent with only moderate modifications. If we do not limit our design within the framework of 802.11, there could be other alternatives with potentially better performance.

\label{}



%% References


\bibliographystyle{elsarticle-num}



\end{document}